# TUNE DETERMINATION OF STRONGLY COUPLED BETATRON OSCILLATIONS IN A FAST-RAMPING SYNCHROTRON*

Y. Alexahin[#], E. Gianfelice-Wendt, W. Marsh, K. Triplett, FNAL, Batavia, IL 60510, U.S.A.

*Abstract*

Tune identification - i.e. attribution of the spectral peak to a particular normal mode of oscillations - can present a significant difficulty in the presence of strong transverse coupling when the normal mode with a lower damping rate dominates spectra of Turn-by-Turn oscillations in both planes. The introduced earlier phased sum algorithm [1] helped to recover the weaker normal mode signal from the noise, but by itself proved to be insufficient for automatic peak identification in the case of close phase advance distribution in both planes. To resolve this difficulty we modified the algorithm by taking and analyzing Turn-by-Turn data for two different ramps with the beam oscillation excited in each plane in turn. Comparison of relative amplitudes of Fourier components allows for correct automatic tune identification. The proposed algorithm was implemented in the Fermilab Booster B38 console application and successfully used for tune, coupling and chromaticity measurements.

## INTRODUCTION

Turn-by-turn (TBT) position measurements of oscillating beam can provide a wealth of information on the machine and beam parameters, such as tunes and optics functions, coupling and chromaticity. The starting point is finding the oscillation tunes.

There is a number of reasons making the betatron tune evaluation from the TBT Beam Position Monitor (BPM) data particularly difficult in fast ramping synchrotrons like the Fermilab Booster: rapid changes of the tunes (sometimes in the course of a few dozen turns), high level of noise, complicate spectra. In the following Section we discuss methods specifically developed for tune evaluation in such conditions: first we recall the phased sum algorithm [1] that made it possible to recover a weak signal from noise, and then introduce new algorithm – which we called a ``two-ramps'' algorithm – for automatic spectral peak identification.

Next we discuss implementation of these algorithms in the Fermilab Booster B38 console application and finally give examples of its usage for tune, coupling and chromaticity measurements.

## BASIC THEORY

*Continuous Fourier Transform*

With a small number of turns $N$ available the resolution of the standard FFT ($1/N$) may be insufficient.

Much better precision can be achieved with the so-called Continuous Fourier Transform (CFT) method [2] which consists in evaluation of the sum

$$X(\nu) = \frac{1}{N}\sum_{n=1}^{N} e^{-2\pi i \nu \cdot n} x_n \qquad (1)$$

as a function of continuous variable $\nu$ and finding the maximum of $|X(\nu)|$.

In absence of random noise CFT provides precision $\sim 1/N^2$, while in its presence the precision is $\sim 1/N^{3/2}$ [1].

An alternative method – the Least Square Fit (LSF) of data with a number of sinusoids (two in our case) has the same tune error dependence on the number of turns [3].

*BPM Phasing*

If the betatron phase advance between BPMs does not differ too much from theoretical values $\varphi^{(k)}_{x,y}$, $k$ being the BPM index, we may try to use information from all available BPMs constructing a phased sum

$$\tilde{x}_n = \sum_k w_k x_n^{(k)} \exp[-i\varphi_x^{(k)}] \qquad (2)$$

for each turn $n$, where $w_k$ are some weights, and perform the CFT analysis using $\tilde{x}_n$ [1]. Weights $w_k$ may reflect the quality of individual BPM data, here we set $w_k=1$.

It is easy to verify that the proper part of the signal propagating with expected phase advance is amplified by a factor of $N_{BPM}$ whereas the alien modes and random noise are amplified only as $\sqrt{N_{BPM}}$ so that the signal to noise ratio is improved by a factor of $\sqrt{N_{BPM}}$ [1]. A more rigorous tune error estimate is:

$$\sigma_\nu \approx \frac{\sqrt{6}\sigma}{\pi a N^{3/2} N_{BPM}^{1/2}}, \qquad (3)$$

where $\sigma$ is the r.m.s. value of BPM errors and $a$ is the betatron oscillations amplitude.

*Two-ramps Algorithm*

In the presence of coupling both transverse mode peaks are present in the spectra of oscillations. The mode with lowest damping rate (e.g. due to low chromaticity) may dominate the spectra of oscillations in both planes making it difficult to attribute spectral peaks to particular modes.

To resolve this problem we propose the following algorithm:

• Take data for two ramps, one with horizontal pings on the beam, the other with vertical pings.

• Looking at the horizontal BPM data from a ramp with h-pings:

▸ Find the highest spectral peak, determine the corresponding tune $Q_{x1}$ and the damping constant, subtract the peak from the data (see Appendix).

▸ Find the next highest peak tune $Q_{x2}$.

---
* Work supported by Fermi Research Alliance, LLC under Contract DE-AC02-07CH11359 with the U.S. DOE.
[#] alexahin@fnal.gov

- Do the same for the vertical BPM data from v-pings to find tunes of the highest and next highest peaks $Q_{y1}$, $Q_{y2}$.
- If the main peaks are well separated, $|Q_{x1} - Q_{y1}| > \varepsilon$, where $\varepsilon \sim 1/N$, then simply: $Q_x = Q_{x1}$, $Q_y = Q_{y1}$.
- If they coincide, $|Q_{x1} - Q_{y1}| < \varepsilon$, then compare the ratios of secondary to primary peak heights in both spectra:
  ➤ If $|X(Q_{x2})/X(Q_{x1})| > |Y(Q_{y2})/Y(Q_{y1})|$ then $Q_x = Q_{x2}$, $Q_y = Q_{y1} \approx Q_{x1}$.
  ➤ If $|X(Q_{x2})/X(Q_{x1})| < |Y(Q_{y2})/Y(Q_{y1})|$ then $Q_x = Q_{x1} \approx Q_{y1}$, $Q_y = Q_{y2}$.

To achieve a better precision in determination of secondary peaks the primary peaks can be subtracted from the spectra as discussed in Appendix. Alternatively, all peaks can be found with a simultaneous least-square fit which gives even better precision but is more difficult to implement in a console application.

Figure 1 gives an example of the spectrum of coupled oscillations of the Fermilab Booster beam which required the application of the described algorithm.

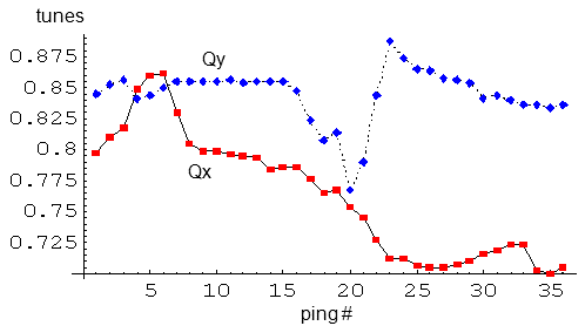

Figure 1 (color): Betatron tunes over the Booster ramp obtained with the help of BPM phasing and two-ramps algorithm. The beam is pinged every 500 turns, i.e. about every 0.8 ms.

## B38 CONSOLE APPLICATION

The Two-Ramps algorithm was implemented in the B38 ACNET control system application making it a truly on-line instrument. This application manages the read out and analysis of the Booster BPM turn by turn data allowing for express measurement of tunes, coupling [4] and – the latest addition – chromaticity. The user may also plot intermediate results such as the BPM data with or without running average subtraction, the phased combined BPM data with or without running average subtraction, and spectral data in the form of mountain range or contour plots. The main user options are shown on the application front page (Fig. 2).

As usual, the chromaticity is measured from the tune dependence on beam momentum. This part of the application works as follows. A kicker, horizontal or vertical, is set up to kick the beam every 500 turns. On completion of the ramp the application reads out the turn by turn BPM data for all turns and all BPMs. Both horizontally pinged data and vertically pinged data are acquired. Pings are identified by detecting the onset of the oscillations.

For each turn the horizontal and vertical BPM readings are separately combined according to equation (2) and for each ping horizontal and vertical CFT's are computed. The tunes are calculated using the above described two-ramps algorithm. Thus the beam momentum is varied by shifting the closed orbit through the radial RF phase feedback. Another set of horizontal and vertical pinged data is acquired. Resorting to the theoretical dispersion the beam momentum relative variation, $\delta=\Delta p/p$, is computed from the closed orbit differences measured at the BPM's and chromaticity, $\xi_{x,y}$, is obtained by the usual expression $\xi_{x,y}=\Delta Q_{x,y}/\delta$.

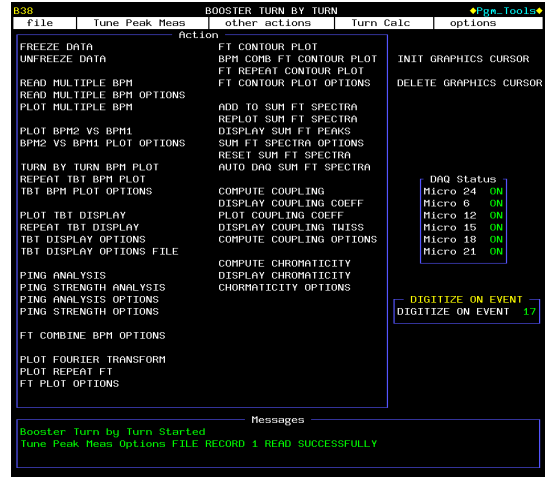

Figure 2: B38 front page.

## RECENT BOOSTER PROGRESS

The new version of B38 application made it possible to calibrate the skew quadrupole and sextupole circuits and correct coupling and chromaticity.

As concerns transverse coupling our goal was to correct the global coupling coefficient $C_-$ [4] through the whole ramp. The absolute value of this coefficient determines the closest tune approach which should be minimized for chromaticity measurements.

The Booster has 24 long and 24 short straights with large vertical and horizontal β-functions respectively. Correctors in the long and short straights can be used as two families of each kind – quadrupole, skew-quadrupole and sextupole – for global correction of the tunes, coupling and chromaticity respectively. The response to the skew-quad and sextupole families was found to be:

$\Delta \text{Re} C_- = (\ 0.0645 \cdot \Delta I_{SQS} + 0.0600 \cdot \Delta I_{SQL}) \cdot 3.183/B\rho$
$\Delta \text{Im} C_- = (-0.0137 \cdot \Delta I_{SQS} + 0.0037 \cdot \Delta I_{SQL}) \cdot 3.183/B\rho$

$\Delta Q_x' = \ \ (8.0 \cdot \Delta I_{SXS} + 0.9 \cdot \Delta I_{SXL}) \cdot 3.183/B\rho$
$\Delta Q_y' = - (0.1 \cdot \Delta I_{SXS} + 2.6 \cdot \Delta I_{SXL}) \cdot 3.183/B\rho$

with currents in Amperes and $B\rho$ in T·m.

With known calibration constants it was possible to reduce coupling to insignificant level in just two iterations. Figure 3 shows the values of coupling coefficient before and after correction.

It is interesting to note that after coupling correction the beam became prone to coherent instability which could have been suppressed by either moving tunes closer to each other (as shown in Fig.1) or increasing chromaticity. More information on the observed instability is presented at this Conference in a different paper [5].

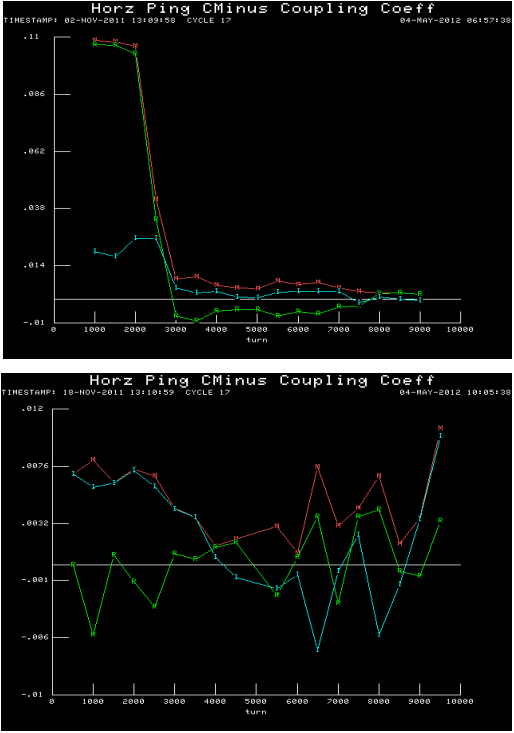

Figure 3 (color): Coupling coefficient over the first part of the Booster ramp before (top) and after (bottom) correction.

Starting chromaticities through the Booster ramp with a large portion around transition energy excluded (pings #12 to #21) are shown in Fig.4 upper plot, while the lower plot presents the final values achieved after a number of studies and corrections (again some region around transition excluded).

## REFERENCES

[1] Y. Alexahin, E. Gianfelice-Wendt, W. Marsh, Proc. IPAC'10, Kyoto, May 2010, p. 1179.
[2] R. Bartolini et al., Particle Accelerators, **56**, 167-199 (1996).
[3] V. Lebedev, private communication.
[4] W. Marsh, Y. Alexahin, E. Gianfelice-Wendt, Proc. PAC'11 New York, March 2011, p. 516.
[5] Y. Alexahin, N. Eddy, E. Gianfelice-Wendt *et al.*, this Conference, WEPPR085.

## APPENDIX

To subtract the main peak from the spectrum we must determine its tune, amplitude, phase and damping constant.
Consider a model signal

$$u_n = e^{2\pi i \nu_0 n}, \quad n = 1,...,N, \quad (A.1)$$

with tune $\nu_0$ possibly complex. Its CFT is $U(\nu) = F[\pi(\nu-\nu_0)]$ where

$$F(x) = e^{-i(N+1)x} \frac{\sin Nx}{N \sin x} \quad (A.2)$$

$$\approx 1 - i(N+1)x - \frac{1}{3}(N+1)(2N+1)x^2 + ...$$

For complex $x = x' + ix''$, $|x| \ll 1$, $|F(x)|$ has a maximum at $x' = 0$ (or $\nu = \text{Re}\,\nu_0$). Information on $\text{Im}\,\nu_0$ can be obtained from the derivative

$$\left.\frac{\partial F}{F \partial x'}\right|_{x'=0} = -i(N+1)(1 + \frac{N-1}{3}x'') + ... \quad (A.3)$$

Now, given the (combined) signal $\tilde{x}_n$, we can find the tune as the position of the maximum of $\tilde{X}(\nu)$ and the damping constant, i.e. the imaginary part of $\nu_0$, from

$$\text{Im}\,\nu_0 = \frac{3}{\pi(N-1)}[1 + \frac{1}{\pi(N+1)} \text{Im} \frac{\partial \log \tilde{X}}{\partial \nu}]_{\nu=\text{Re}\,\nu_0} \quad (A.4)$$

Subtraction of the found main peak is performed as

$$\tilde{\tilde{x}} = \tilde{x} - u \frac{<u, \tilde{x}>}{<u, u>}, \quad <w, u> \equiv \frac{1}{N}\sum_{n=1}^{N} w_n^* u_n, \quad (A.5)$$

and does not explicitly involve the amplitude and phase of TBT oscillations.

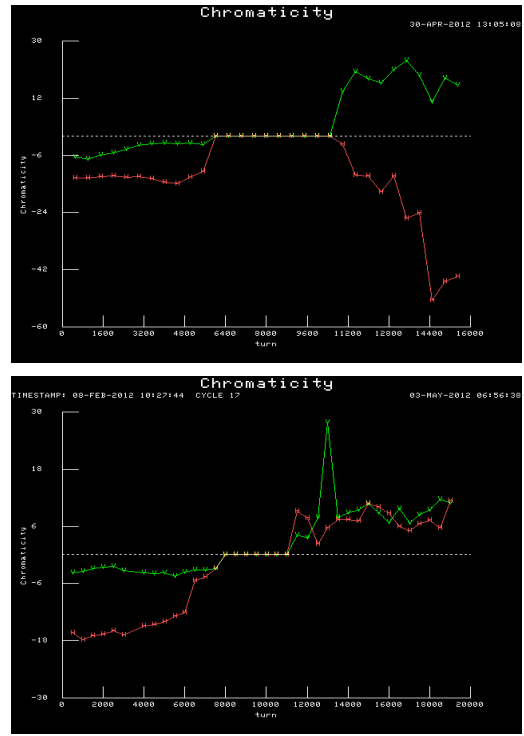

Figure 4 (color): Chromaticities as found 11/17/2011 (top) and after 02/08/2012 studies (bottom) with the central part of the ramp excluded around transition energy excluded. The spike in $Q_y'$ in the lower plot is an artefact of measurement.